\documentclass[letterpaper,sort&compress,5p]{elsarticle}

\usepackage{graphics,braket,amsmath,color}
\usepackage[usenames,dvipsnames,svgnames,table]{xcolor}

\journal{Solid State Communications}

\title{Unexpectedly Slow Two Particle Decay of Ultra-Dense Excitons in Cuprous Oxide}

\author[nu]{N. Laszlo Frazer\corref{cor}}
\ead{ssc@laszlofrazer.com}
\address[nu]{Department of Physics, Northwestern University, 2145 Sheridan Road Evanston, IL 60208-3112 USA}
\cortext[cor]{Corresponding author. 1(773)634-9638}

\author[cnm]{Richard D. Schaller}
\ead{schaller@anl.gov}

\address[cnm]{Center For Nanoscale Materials, Argonne National Laboratory and Department of Chemistry, Northwestern University, Building 440, 9700 South Cass Avenue, 
 Argonne, IL 60439 USA}

\author[nu]{J. B. Ketterson}
\ead{j-ketterson@northwestern.edu}

\date{\today}

\begin{document}

\begin{abstract}
	For an ultra-dense exciton gas in cuprous oxide (Cu\begin{math}_2\end{math}O), exciton-exciton interactions are the dominant cause of exciton decay.  This study demonstrates that the accepted Auger recombination model overestimates the exciton decay rate following intense two photon excitation.  Two exciton decay is relevant to the search for collective quantum behavior of excitons in bulk systems.  These results suggest the existence of a new high density regime of exciton behavior.
\end{abstract}

\maketitle

\section{Introduction}

Since quantum mechanics is generally viewed as a microscopic theory, macroscopic quantum phenomena (MQP) such as superconductivity, superfluidity, and Bose-Einstein Condensation (BEC) \cite{landau1941theory,greiner2002quantum,onnes1911resistance,bardeen1957theory,bose1924plancks,anderson1995observation} have had a dramatic impact.  Wannier-Mott excitons are an excellent system to search for novel MQP due to their low mass, solid-state environment, and Rydberg-like structure \cite{apfel1955exciton}.  Indeed, two dimensional exciton-polariton condensation has been observed \cite{deng2002condensation,timofeev2012exciton}. 

Cuprous oxide is an ideal material for probing three dimensional bosonic MQP \cite{mysyrowicz1979long} and there have been both reports \cite{lin1993bose,fortin1993exciton,benson1997anomalous} and reinterpretations \cite{tikho,warren2000two,tikhodeev2000exciton,snoke2002spontaneous} of BEC and superfluidity.  Recently, it has been proposed that an observed relaxation explosion indicates the generation of a BEC \cite{yoshioka2011transition}.  Low temperature, low mass, long lifetimes, and high density all contribute to the formation of MQP \cite{schwartz2012dynamics}.  Of these variables, the density has received the least attention.  In this paper excitons are excited using an intensity approximately \begin{math}10^6\end{math} times higher than that used in previous studies \cite{yoshioka2011transition,yoshioka2006dark}. The dense exciton lifetimes exceed predictions based on a configuration independent exciton-exciton annihilation model and suggest that for this system, which is selectively excited, the decay rate falls below the Auger predictions.

 Nonradiative decay of exciton pairs at high densities creates a difficulty; while MQP have a higher transition temperature in a denser gas, higher density is accompanied by a shorter lifetime and heating.  In fact, it has been reported that exciton lifetimes are inversely related to exciton density \cite{kavoulakis1996auger,kavoulakis2000auger,ohara1999auger,warren2000two}.  Here we present evidence that exciton lifetimes are longer following the sudden creation of a high density population.  Since there is insufficient time for coherence to develop spontaneously, thermalization mechanisms applicable to dilute excitons \cite{yoshioka2011transition,warren2000two} should not apply at high densities. However, since the observed decay rates of dense excitons are smaller than expected, it may be possible to non-spontaneously generate MQP in the future.  An analysis of exciton production, state mixing, and decay processes shows that particle interactions play a greater role in intensely excited systems.

\section{Methods}

\begin{figure*}
	\begin{center}
\includegraphics{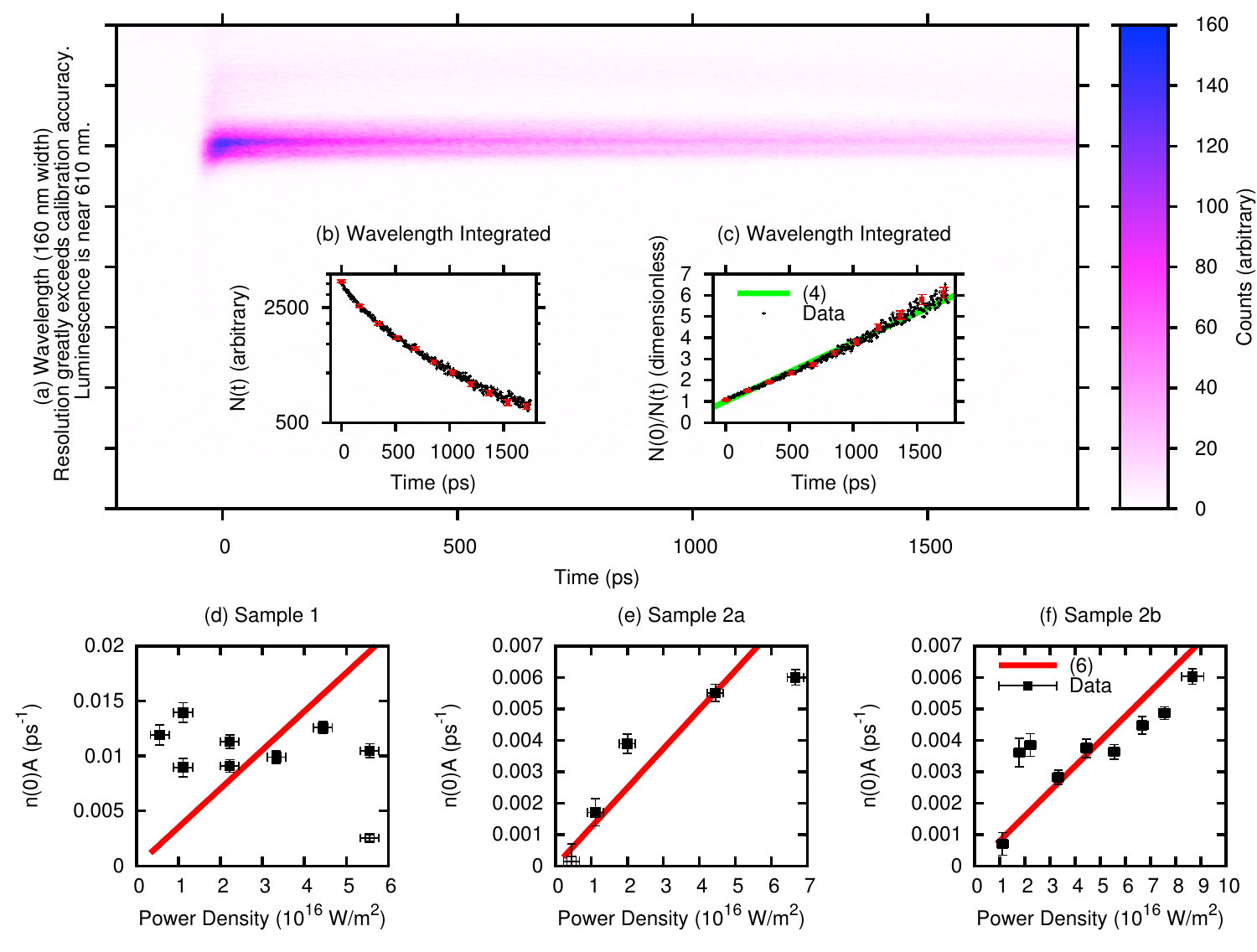}
	\end{center}
\caption{
(a) Time and wavelength resolved exciton luminescence from Sample 2 excited at $ 3.3\cdot 10^{16}$ W/m$^2$, averaged over $6\cdot 10^7$  laser shots.  For absolute wavelength, see the supporting information. The data integrated over wavelength plotted on a log linear scale to show the decay is not exponential (b) and reciprocal linear scale to show that two body decay dominates (c).  Representative uncertainties are shown.  (d-f) The initial decay rate of a dense exciton gas $ n(0)A $ as a function of the peak power density of the excitation laser showing discrepancies from Equation (\ref{EM}). For Sample 2, two different excitation locations are shown. Three statistical outliers excluded from the analysis have open symbols; the two outliers for Sample 1 are on top of each other. The third outlier occurred because insufficient photons were collected at the lowest powers. The data in (a) was collected from a separate location with greater averaging. Only systematically collected data were included in the analysis.}
\label{average}
\end{figure*}

\begin{figure*}
	\begin{center}
	\includegraphics{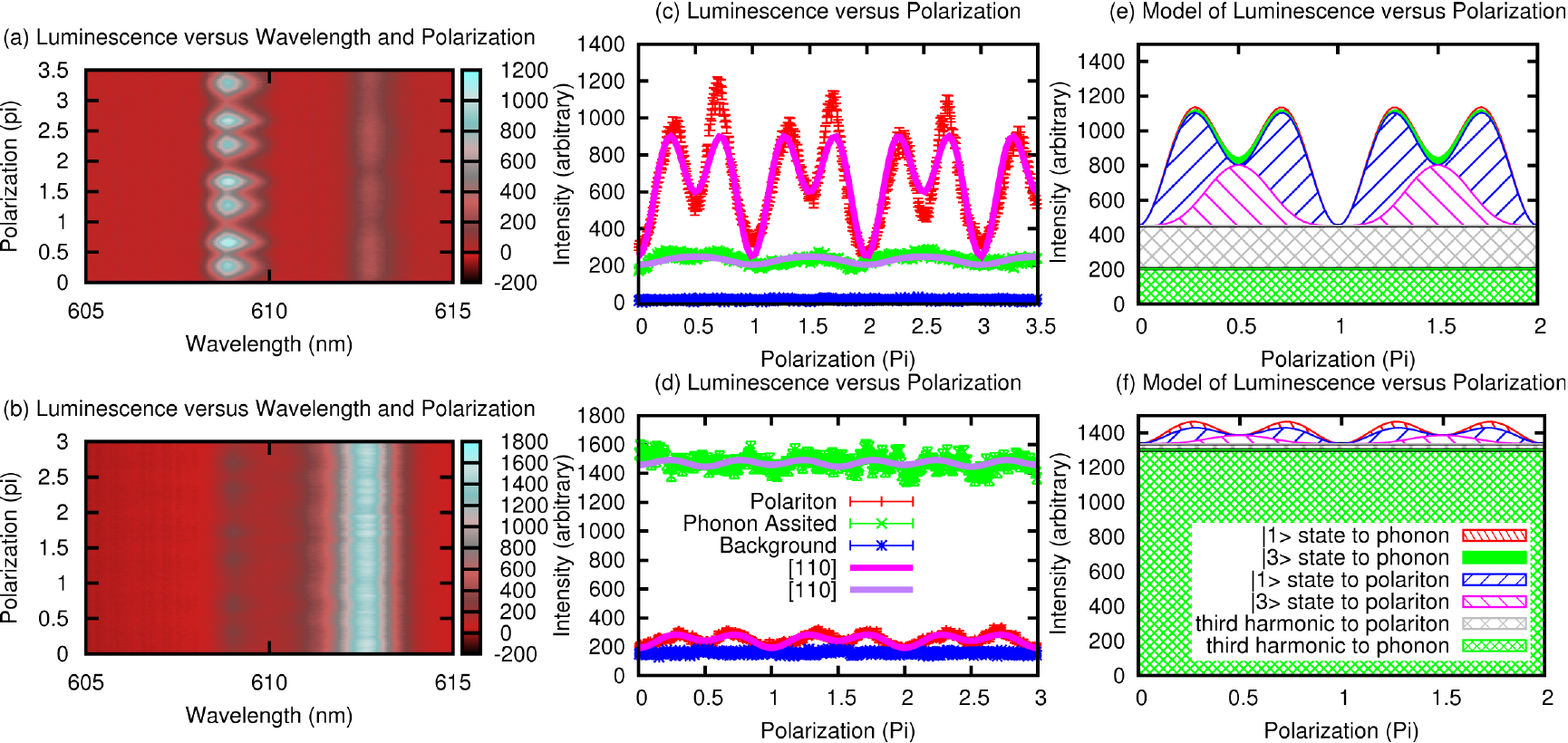}
	\end{center}
\caption{\label{polar} 
(a, b) Time integrated exciton luminescence spectrum from Sample 2 versus angle between the pump electric field and the $ \left[ 100 \right]$ The left peak is the polariton decay process.  The right peak is the phonon assisted decay process.  (c, d)  Luminescence from the two peaks fitted with the sum of the selection rules.  (e, f) Model showing the cumulative brightness that would be observed by the streak camera broken down by production and decay channel.   (a, c, e) $ 0.4\cdot 10^{16}$ {W}/{m}$^2$ excitation.  (b, d, f) $ 4.7\cdot 10^{16}$ {W}/{m}$^2$ excitation.  In (f), the $\ket3$ to phonon assisted decay process is not shown since it was not statistically significant.
}
\end{figure*}

Our excitation scheme \cite{ideguchi2008coherent,inoue1965two} is comparable to the one used in Ref. \cite{yoshioka2006dark} to produce low density excitons by two photon excitation and time resolve their decay luminescence (Figure \ref{average} (a)). The method allows the production of excitons which all have essentially the same initial momentum and energy corresponding to luminescence with a width of about 0.001 nm out of 610 nm \cite{fr1991coherent}, and avoids the direct creation of electron-hole pairs.  However, in this experiment, where the excitation intensities are 
\begin{math}
	10^{6}
\end{math} 
 times higher than previous work (Figure \ref{average} (d-f)), the pump wavelength is on resonance for the excitation of both ortho- and the para-excitons \cite{liu2005resonant}, and the pump laser has variable polarization (Figure \ref{polar}).  Except where noted, the polarization was selected to produce the maximum number of excitons in accordance with the selection rules for normal incidence on [100] (Sample 1) and [110] (Sample 2) faces.  In addition, a spectrometer was used to simultaneously resolve the wavelength {\em and} arrival time of the transmitted signal using a streak camera (Figure \ref{average} (a)).   The spectrograph entrance slit was similar in size to the excitation spot image, so low density excitons, which may have moved away from the spot, are largely excluded from the data.  The streak camera used has ten times higher time resolution than Ref. \cite{yoshioka2006dark}.  Time averaged spectra were also collected (Figure \ref{1400} and in the supplement).  Details of the methods are in the supplement.

\section{Results and Discussion}
\subsection{Two exciton decay process}

Cuprous oxide is a semiconductor with a band parity such that radiative decay of excitons is either suppressed or forbidden. Suppressed cases include the polariton forming exciton state, decays involving both a phonon and a single photon, and perturbations arising from strain \cite{liu2005resonant} or electric and magnetic fields.  The polariton forming state occurs when the lowest energy orthoexciton (which has spin one) matches the photon momentum. On exiting the crystal it reverts back to light.  While propagating, the exciton polariton is a quantum superposition of a photon and an exciton.  The polariton is especially attractive since the photon component reduces exciton interactions and its effective mass.  The effective mass follows from the local curvature of the dispersion in the crossover region \cite{fr1991coherent}.  There is no exciton-photon coupling at zero momentum in cuprous oxide due to parity conservation, but the polariton forming state has the advantage that it is easily observed via higher-order coupling to the light field.

Low density excitons primarily undergo a decay that can be modeled by
\begin{align}
	\frac{dn}{dt}=-\frac{n(t)}{\tau}\label{lowdense}
\end{align}
where \begin{math}
	n
\end{math} is the exciton density and \begin{math}\tau\end{math} is the relaxation time, which has been to reported to be up to \begin{math}13\end{math} \begin{math} \mu\end{math}s \cite{shen1997dynamics,kubouchi2005study,mysyrowicz1979long,yoshioka2006dark,sun2001production}.

	Two excitons can undergo an ``Auger'' reaction where one electron and one hole annihilate, imparting their energy to the remaining unbound electron and hole which are short lived.  Most cuprous oxide studies show no spectroscopic evidence of free carriers because they decay nonradiatively, but one report shows it at the \begin{math}10^{-3}\end{math} level \cite{oharadissertation}.  If the two exciton process is dominant in a dense exciton gas, the detection rate, as observed through the polariton and phonon assisted decay processes, will behave as

\begin{align}
	\frac{dn}{dt}=-An(t)^2\label{vague}
\end{align}
In low-density experiments \cite{ohara1999auger,warren2000two,jolk2002exciton,jang2006exciton,jang2006auger,jang2008suppression,mani2010nonlinear} the Auger Constant \begin{math}
	A
\end{math} is independent of the details of the excitation mechanism.  Variations in the configuration of the excitons are not considered in the model.  It has often been applied to lower intensity pulsed experiments, where variations in the momentum of the excitons were not evident in the measured Auger constant.  The literature provides a wide range of values for \begin{math}A\end{math}  extending over four orders of magnitude \cite{jolk2002exciton,ohara1999auger,warren2000two,denev2002stress,jang2006auger,schwartz2012dynamics,yoshioka2010quantum,jang2008suppression,mani2010nonlinear,kavoulakis1996auger,kavoulakis2000auger,jang2005biexcitons,jang2006exciton}  including configuration variations due to decay, expansion, or temperature.  Integrating Equation (\ref{vague}) yields

\begin{align}
	n(t)&=\frac{n(0)}{An(0)t+1}&t\ge 0\label{twobody}
\end{align}
The density \begin{math} n \end{math} is difficult to calibrate \cite{jolk2002exciton,ohara1999auger,warren2000two,denev2002stress,jang2006auger,schwartz2012dynamics,yoshioka2010quantum,jang2008suppression,mani2010nonlinear,kavoulakis1996auger,kavoulakis2000auger,jang2005biexcitons,jang2006exciton}  and absolute measurements were therefore not attempted. From an experimental point of view the value \begin{math}An(0)\end{math} is useful because it is the typical decay rate of excitons at their peak density.  This experiment is designed to measure \begin{math}An(0)\end{math}.  To relate (\ref{twobody}) to Figure \ref{average} (d-f), it is convenient to rewrite it as
	\begin{align}
		\frac{n(0)}{n(t)}&=An(0)t+1&t\ge 0 \label{rewrite}
	\end{align}
	where we see that the slope of Figure \ref{average} (c) is the vertical axis of Figure 1 (d-f).

	\subsection{Evidence for density dependence of decay rate}
	Based on the 2 ns dynamic range of the streak camera and our experience from previous experiments \cite{ohara1999auger,warren2000two,jang2006auger}, only the transient, high density behavior can be tracked.  The wavelength-integrated number of photons from exciton radiative decay detected in each time bin \begin{math} N(t) \end{math}  is used as a measure of the relative instantaneous exciton density: \begin{math}
		N(t)\propto n(t)
	\end{math}.  As in most optical detection experiments, there is an assumption that the excitons which radiatively decay are a reasonable sample of the overall exciton population.  To characterize the decay, Equation (\ref{twobody}) was fitted to the wavelength-integrated streak camera to determine \begin{math}
		N(0)
	\end{math} and \begin{math}
		An(0)
	\end{math}  The proportionality constant relating \begin{math}
		N(t)
	\end{math} and \begin{math}
		n(t)
	\end{math} is not determined and is not needed to obtain \begin{math}
		An(0)
	\end{math} as long as proportionality is valid. The onset point \begin{math}
		t=0
	\end{math}, which is not fit as part of a regression, was picked to be the time bin at which the largest number of photons arrives.  It is very close to the time at which the laser pulse arrives.

	The fit of (\ref{twobody}) was compared to the density independent exciton decay model (Equation (\ref{lowdense})):
	\begin{align}
		n(t)&=n(0)e^{-\frac{t}{\tau}}&t\ge 0
		\label{onebody}
	\end{align}
	Low density excitons follow this behavior. The behavior in Figure \ref{average} (b) is nonlinear while that in \ref{average} (c) is linear, from which we conclude that two exciton decay dominates one exciton decay.  Since the density independent model includes exciton decay at an impurity, it is apparent from dynamics as well as spectra that impurity effects are not important here.

	A sum of one and two body decay processes was not considered for three reasons:  First, if the low density exciton decay influenced the observed dynamics, it would cause underestimates of the two exciton decay rate at lower excitation powers.  These rates are higher than predicted by the model.  Second, if $\tau$ is the literature value \cite{yoshioka2006dark}, then insufficient photons were collected to be able to measure it from the low density tail of the streak camera data, especially considering the temporal dynamic range of the streak camera.  Third, attempts to measure \begin{math} An(0) \end{math} and \begin{math} \tau \end{math} by fitting simultaneously were frustrated by their strong covariance.  We conclude that the data are in a regime where only multi-exciton processes are important to population changes.  This explains the variation in the dynamical data very well.

The observed dynamics verify that high excitation intensities lead to strong exciton interactions.  The method avoids determination of the absolute exciton density, which is subject to challenging systematics.

		\subsection{Intense excitation leads to an effective reduction in the Auger constant}
		Using a constant \begin{math} A \end{math} and given a large peak laser intensity \begin{math} P \end{math}, the initial decay rate model is \cite{mani2010nonlinear}
	\begin{align}
		An(0)&\propto P
	\label{EM}
	\end{align}
It is expected to be valid when the exciton distribution function is not evolving, and possibly other times as well.

	To highlight deviations from this relation, the value of \begin{math} P \end{math}  was made very large.  The experiment cannot decouple \begin{math} A \end{math} and \begin{math} n(0) \end{math}, but \begin{math} n(0)\propto P \end{math} is consistent with the data (supporting information) and expectations \cite{mani2010nonlinear}.

Previous work was oriented towards evolving the excitons towards a Bose-Einstein distribution function.  In this experiment, we maximize occupation of the momentum state produced by two photon absorption.  The initial distribution function has a delta-function like shape along the momentum axis, which is made as big as possible.  This should reduce interactions since many of the excitons have the same momentum, leading to a smaller \begin{math} An(0) \end{math} than would otherwise occur.  We will see that additional components to the distribution function become increasingly important as excitation becomes more intense.  Changes in the initial distribution function with intensity cause variations from the Auger constant associated with the Bose-Einstein distribution.  Uneven evolution of the distribution function should lead to a time evolving Auger constant, but apparently this is not strong enough to detect.

	Nonlinear optical processes can become less efficient at high intensities.  Equation (\ref{EM})  already includes suppression of the net production of excitons via a two photon process from quadratic to linear possibly due to exciton-exciton annihilation.  Additional suppression mechanisms could lead to further changes towards smaller \begin{math} An(0) \end{math} at higher intensities.  For example, electron-hole plasma could reflect the pump beam and decay nonradiatively.  However, the data does not show a convincing deviation from \begin{math} n(0)\propto P \end{math} (supporting information).

		In Figure \ref{average} (d-f), the decay rate model with an Auger constant is statistically rejected with high confidence in this dense exciton experiment (supporting information).  If it is assumed that the laser power is not high, or that three photon excitation is the primary source of excitions, then new models form which deviate more strongly from the data.

\subsection{Polarization}
The orthoexciton state can be described by a basis of three degenerate spin states \begin{math}\left\{ \ket1,\ket2,\ket3 \right\}\end{math} in the notation of Ref. \cite{yoshioka2006dark}.  The two-photon polarization selection rules \cite{elliott1961symmetry,inoue1965two}  allow the production of the \begin{math}\ket{1}\end{math} and \begin{math}\ket{3}\end{math} states for propagation along the \begin{math} \left[ 110 \right] \end{math} crystal axis; two-photon excitation of the \begin{math} \ket2
\end{math} state is forbidden.  
The dependence on polarization for two different excitation powers is shown in Figure \ref{polar}.  
Note there is a component of the luminescence that is not consistent with the selection rules.  The laser beam was carefully filtered to ensure the absence of any components over the band gap that might result in a polarization independent fraction.  Therefore, the polarization independent features have been assigned to absorption of third harmonic photons to make electrons and holes which later become excitons \cite{mani2010nonlinear,liu2005resonant}.   In fact, intense 407 nm luminescence from third harmonic generation in cuprous oxide was observed with the streak camera.  A three photon process is consistent with a polarization independent exciton production mechanism overwhelming the polarization dependent process at high powers.  It is also consistent with most of the polarization independent excitons decaying through the phonon assisted channel, which is not sensitive to the exciton momentum.

\begin{figure}

	\includegraphics{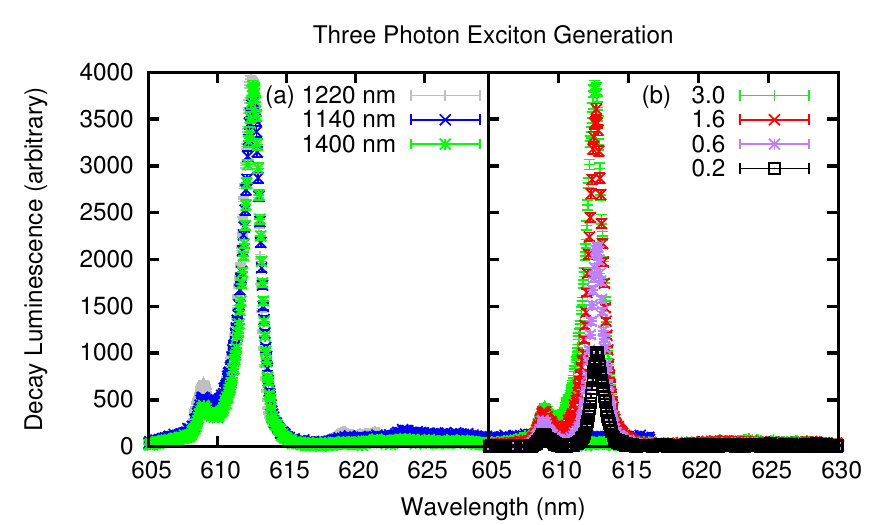}

\caption{\label{1400} 
Time averaged exciton luminescence spectra of Sample 2 (a) for a pump power of  $3\cdot10^{16}$ W/m$^2$ with indicated excitation wavelengths and (b) for 1400 nm excitation, with indicated pump powers in units of $10^{16}$ W/m$^2$.
 }
\end{figure}

Since third harmonic generation is not specific to a particular photon energy, any resulting excitons should be present when the pump laser is tuned to other wavelengths.  At high illumination intensities, changing the wavelength of the laser had no large effect on the exciton spectrum (Figure \ref{1400}) or lifetime.  At lower illumination intensities, strong polariton decay luminescence was not observed off the two photon resonant energy, though it is sometimes observed with resonant excitation (see supporting information) since two-photon excitation can directly create excitons.

\subsection{Luminescence Rise Time}
The two-photon selection rules in \begin{math} \left[ 110 \right] \end{math} cuprous oxide only permit the production of excitons that do not form polaritons.  The reported time for low density excitons to convert to a polariton forming state is \begin{math} 244\pm2 \end{math} ps \cite{yoshioka2006dark}.  Here, no conversion time was observed for times as short as 2.5 ps.  If the conversion between exciton states is due to exciton-exciton interactions, the expected conversion time would be extremely short.  Rapid state mixing supports the assumption that radiative decay samples the exciton density in a representative manner

\section{Conclusions}

\begin{figure*}
	\begin{center}
	  \includegraphics[scale=.9]{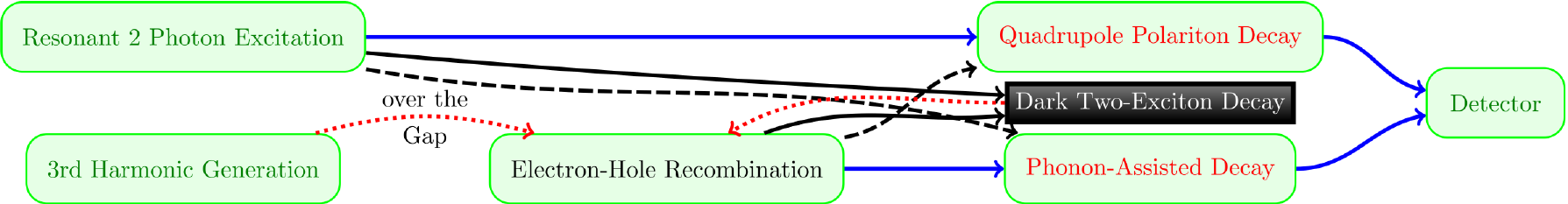}
	\end{center}

\caption{\label{diagram} 
Flow chart for establishing that two-exciton decay is slower than expected.  Dotted paths are for electrons and holes.  Dashed paths are less likely due to momentum conservation.  Impurity effects are not shown.
}
\end{figure*}

A diagram of the processes studied in this experiment is shown in Figure \ref{diagram}.  Polarization analysis, off resonant wavelength excitation, and direct observation of harmonics indicate that a portion of the exciton gas is created by third-harmonic over-the-gap light generated within the crystal.  In addition, the observation that the rise time of the exciton luminescence is very short suggests that different exciton states mix rapidly when the exciton gas is dense.

Under intense excitation, the well known but not previously well tested model that the two exciton decay rate is proportional to the excitation power is strongly excluded by our measurements of the exciton luminescence dynamics.  Some of the data are consistent with a lower limit on the exciton lifetime of around \begin{math} 92\pm 5 \end{math} ps, but in no case did the dense exciton gas have a characteristic time less than 70 ps, despite reports that under typical experimental conditions that time is approximately 13 ps \cite{warren2000two}.  The observed lifetimes are \begin{math}10^3\end{math} longer than the inferred trapping time of cold paraexcitons undergoing a relaxation explosion \cite{yoshioka2011transition}.  For all of the data sets, an Auger constant extrapolated from low excitation intensities leads to a gradually increasing overestimate of the observed decay rate at higher intensities.

	Each of the lines of evidence suggest that new interactions develop at high excitation intensities, which might include a change from Bose to Fermi statistics \cite{greiner} which can block bosonic MQP \cite{snoke2012polariton}.  It is crucial to develop an understanding of the impact of those interactions on the density threshold of MQP.

\section{Acknowledgement}
We would like to thank Professor M. Grayson for helpful discussions.  This work made use of the J. B. Cohen X-Ray Diffraction Facility and OMM Facility supported by the MRSEC program of the National Science Foundation (DMR-0520513) at the Materials Research Center of Northwestern University.  Use of the Center for Nanoscale Materials was supported by the U. S. Department of Energy, Office of Science, Office of Basic Energy Sciences, under Contract No. DE-AC02-06CH11357.  Support was provided by NSF IGERT DGE-0801685.  N. L. F. gratefully acknowledges support from the Ryan Fellowship and the Northwestern University International Institute for Nanotechnology.


\appendix

\section{Supplementary Data}
Details of methods, statistics, initial intensities,  time averaged spectra, exciton spatial distribution, and raw data are available.

\bibliographystyle{model1a-num-names}
\bibliography{p}
\end{document}


\begin{keyword}A. copper(I) oxide; D. Polaritons; D. Auger; E. femtosecond laser
 \end{keyword}
\maketitle

\section{Experimental Methods Details\label{details}}
A 35 fs pulsed Ti:sapphire laser with a 2 kHz repetition rate was used to pump an optical parametric amplifier.  A very short pulsed laser is necessary to achieve very high excitation intensities.  The amplifier was tuned to a wavelength of 1220 nm with a half width of about 20 nm so that the laser wavelength was centered on the two-photon excitation resonance of the orthoexciton but overlapped the stress-based two-photon resonance of the paraexciton in cuprous oxide \cite{liu2005resonant}.  The breadth of the laser line explains the absence of quantum beats \cite{fr1991coherent} between the polariton dispersion branches.
Two-photon excitation was used so that exciton luminescence could be collected from the transmission side of the sample without the introduction of any background signal from the laser.  In addition, since second harmonic generation is forbidden in cuprous oxide and overlap of the laser spectrum with half the energy of the band gap is very small, no free electrons and holes are expected to be directly produced by two-photon processes.  

The laser beam was focused to a spot of \begin{math} 90\pm18\mu \end{math}m diameter with a 125 mm convex lens.  The spot was positioned in a natural cuprous oxide crystal in a vacuum optical cryostat cooled to about 3.4 K.  The laser heats the sample, but spectra indicate that the sample temperature is much less that 10 K.  Sample 1 is about 3.6 by 2.3 by 1.3 mm and Sample 2 is 2.5 by 1.8 by 1.3 mm.  
		        The large faces of these approximate rectangular prisms are polished (100) and (110) surfaces respectively, which were verified by the Laue X-ray method.  Samples were mounted with little strain on sapphire with the large face normal to the laser beam.

\begin{figure}
	\includegraphics{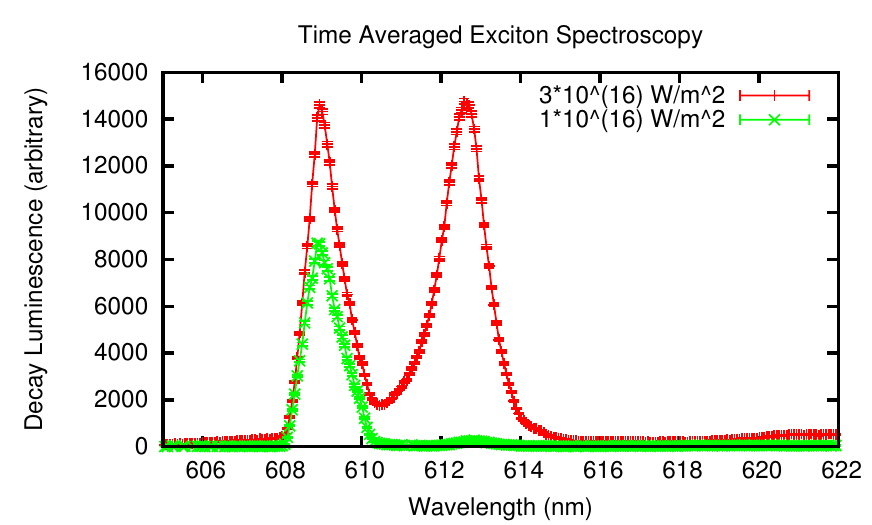}
	\caption{\label{timeaveraged}
	Time averaged spectrum of two-photon excited luminescence in Sample 1 of cuprous oxide with different laser powers. The left peak is the orthoexciton polariton decay process.  The right peak is the phonon assisted decay process.  Radiative decay of the paraexciton is blocked by parity conservation.
	}
\end{figure}

\begin{figure}
	\includegraphics{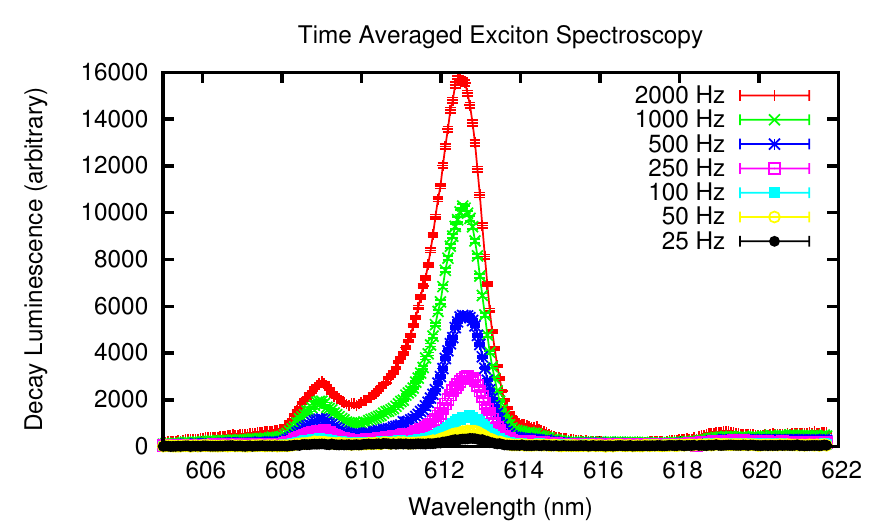}
	\caption{\label{rate}
	Time averaged spectrum of two-photon excited luminescence in Sample 2 of cuprous oxide with various laser pulse rates.
	}
\end{figure}

			        Light was collected in a spectrograph.  At the exit to the spectrograph, a streak camera was used to measure the arrival time of single photons with 22 ps response time and 2 to 5 nm resolution (Ref. \cite{yoshioka2006dark} does not provide any spectral information for their streak camera, which has 40 ps resolution).  The system also had a large systematic spectral error which was sensitive to alignment.  The streak camera's spectral capabilities can be used to exclude third harmonic generation, bound exciton luminescence, and laser light from the data.  Since the streak camera was not consistently able to resolve the phonon-assisted decay and the polariton-assisted decay of excitons separately, luminescence was also collected by a higher resolution, time averaged spectrometer (Figures \ref{timeaveraged} and \ref{rate}).

\section{Statistics}
In Figure 1 (c) of the main text, the coefficient of determination \begin{math}R^{2}=0.998\end{math}.  This indicates that variation in the dynamical data is well described by the Auger model.

\begin{table}
        \begin{center}
\begin{tabular}{l|r|r|r}
&Sample 1&Sample 2a&Sample 2b\\\hline
P-value& {0.002}&{0.01}&{0.0001}
\end{tabular}
        \end{center}
\caption{P-values from reduced $\chi^2$ statistics indicate that discrepancies from the relationship $An(0)\propto P$ are not due to chance.  The data is shown in Figure 1 (d-f).\label{pvalue}}
\end{table}
P-values for relationship DM are shown in Table \ref{pvalue}.    It is important that the p-values were computed using the time and power uncertainties shown in Figure 1 (d-f).  Uncertainty propagation through analysis steps has ensured the data is not biased towards the dynamic range of the streak camera.  In fact, lifetimes similar to the dynamic range of the streak camera have greater uncertainties because they correspond to low density cases where fewer photons were detected.  The most likely cause of differences between samples is systematic variation in the efficiency with which the laser beam is coupled into the sample, with contributions from differing selection rules.

\section{Initial Intensities}
\begin{figure*}
	\begin{center}
\includegraphics{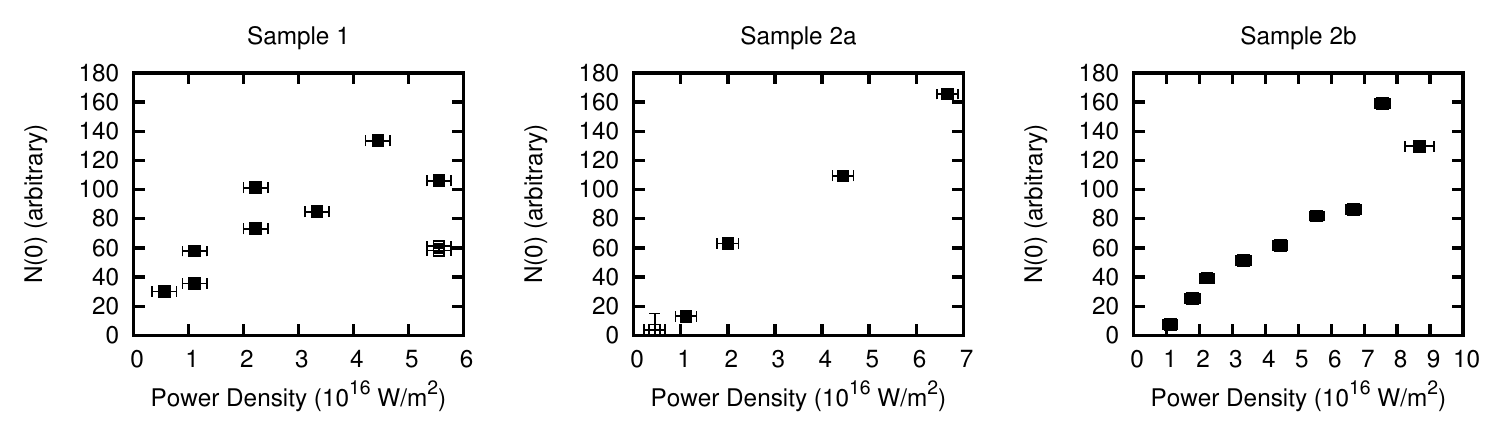}
	\end{center}
\caption{\label{Initial}  Initial luminescence intensities are approximately proportional to the excitation intensity.
	}
\end{figure*}

Initial intensities are proportional to the laser peak power (Figure \ref{Initial}).  Two photon absorption combined with Auger processes can produce this linear behavior.  Detection efficiencies are not the same for each sample.

\section{Time Averaged Spectra Discussion\label{timeaveragedisc}}
No paraexciton line was found due to the absence of stress \cite{liu2005resonant}.  While the phonon assisted decay produces two particles, a phonon and a photon, the polariton assisted decay produces only a photon.  The relative size of the two peaks in Figure \ref{timeaveraged} varies because polariton assisted decay is subject to a strict momentum conservation requirement.  At higher laser intensities, more excitons will scatter so that they no longer have the necessary momentum to form a polariton oriented towards the detector.  In addition, sample quality and optical alignment may influence the relative sizes of the peaks.  Figure \ref{timeaveraged} is from Sample 1 with a 2 kHz laser pulse rate.  Figure \ref{rate} is from Sample 2, under conditions similar to the first figure in the article, but with varying laser pulse rates.  The relative size of the two peaks varies, but is not sensitive to the laser pulse rate.

\section{Spatial Distribution of Excitons}

The Rayleigh length of the focus does not extend over the entire thickness of the sample.  The width of the laser's guassian profile is \begin{math}
	\sigma
\end{math}.  Then the expected initial distribution of two photon produced excitons is a gaussian with width \begin{math}
	\frac{\sigma}{\sqrt2}
\end{math}.  The electrons and holes initially produced by third harmonic generation are distributed in a gaussian with width \begin{math}
	\frac{\sigma}{\sqrt3}
\end{math}.  These distributions are essentially the same.  Electron-hole binding into excitons is too rapid to change the shape of the exciton spatial distribution.

At long times, the shape of the exciton gas evolves diffusively with \begin{math}
	\sigma\sim\sqrt{Dt}
\end{math} and amplitude \begin{math}
	\propto \frac{1}{Dt}
\end{math}.  Since \begin{math}
	D\approx 200
\end{math} cm\begin{math}^2\end{math}/s, in the time scale of 92 ps diffusive expansion is about 1 micron \cite{trauernicht1984highly}, which is negligible.  The impact of repulsive expansion on the spatial distribution of excitons is uncertain.

 \section{Available Supplemental Raw Data}
 Attached is the raw data for

 \begin{itemize}
\item Figure 1 (d) files 1-10
\item Figure 1 (e) files 11-15
\item Figure 1 (f) files 16-24
\item Figure 2 (a) file 25
\item Figure 2 (b) file 26
\end{itemize}

Example metadata is provided.  The time window is 2050 ps. Average laser power is indicated in the file name.  Calibration with a mercury argon lamp shows that the wavelength values in the time averaged spectra are 0.15\begin{math}\pm\end{math}0.01 nm too high.  This is corrected in the figures.

	\bibliographystyle{model1a-num-names}
	\bibliography{p}